\documentclass[doublecol]{epl2}
\usepackage{amsmath}

\title{Identifying the orbital angular momentum of light based on atomic ensembles}
\shorttitle{Identifying the orbital angular momentum of light based on atomic ensembles} 

\author{Liang Han\inst{1} \and Mingtao Cao\inst{1} \and Ruifeng Liu\inst{1} \and Hao Liu\inst{1} \and Wenge Guo\inst{2} \and Dong Wei\inst{1} \and Shaoyan Gao\inst{1} \and Pei Zhang\inst{1}\thanks{E-mail: \email{zhangpei@mail.ustc.edu.cn}} \and Hong Gao\inst{1}\thanks{E-mail: \email{honggao@mail.xjtu.edu.cn}} \and Fuli Li\inst{1}}
\shortauthor{Liang Han \etal}

\institute{
 \inst{1}Department of Applied Physics, Xi'an Jiaotong University, Xi'an 710049, China\\
 \inst{2}School of Science, Xi'an Shiyou University, Xi'an 710065, China
}

\abstract{We propose a scheme to distinguish the orbital angular momentum state of the Laguerre-Gaussian (LG) beam based on the electromagnetically induced transparency modulated by a microwave field in atomic ensembles. We show that the transverse phase variation of a probe beam with the LG mode can be mapped into the spatial intensity distribution due to the change of atomic coherence caused by the microwave. The proposal may provide a useful tool for studying higher-dimensional quantum information based on atomic ensembles.}

\begin{document}
\maketitle
Photon is a promising candidate for quantum information processes. Photons with the Laguerre-Gaussian (LG) mode possess both spin angular momentum (SAM) and orbital angular momentum (OAM)\cite{LA}. The SAM relates to the light polarization, while the OAM relates to the transverse angular phase of light in the form of $\exp (il\varphi )$. The LG beam has attracted much attention in the last two decades\cite{SF,AM}, as OAM states of light can be encoded as the higher-dimensional quantum information \cite{GF}. The higher-dimensional quantum system can be engineered to the quantum repeater based on atomic ensembles \cite{LM}, which plays a crucial role in realizing long distance quantum communication. Recently, there are many works about the propagating of LG modes through atomic ensembles. Such as, generating the entanglement of OAM states of  photon pairs in a hot atomic ensemble \cite{QF}, narrowing the electromagnetically induced transparency (EIT) spectrum linewidth by LG modes\cite{SR}, modulating LG modes through a non-material lens in a vapor cell \cite{DS}, transferring LG modes from control beam to probe beam in atomic gases\cite{JR} and transferring the frequency of LG modes in four-wave mixing\cite{GA}. On the other hand, measuring LG modes is the basic and important problem in the applications of OAM. There are also many works about identifying LG modes by interference\cite{JL,JC}, diffraction\cite{GCG,JM} and image reformatting\cite{GB}. Considering the interesting application of LG modes in the atomic ensembles, it is worth to study the identification of the OAM of light based on atomic ensembles. This may provide us an interface for higher-dimensional quantum information between OAM and atomic ensembles. Here, we propose a scheme to achieve the identification of LG modes in atomic ensembles. This method also has other potential applications in the aspect that based on atomic ensembles. For example, it is possible to achieve efficient light storage and retrieve for complex image. As we know, the image storage and retrieve in atomic ensemble is affected by optical diffusion\cite{MS}, which we always want to diminish\cite{OF}. Our method may be useful for reducing the effect of diffusion on the visibility of the reconstructed image\cite{MS}.

EIT is a quantum interference phenomenon based on the coherent population trapping. Since it was found by Harris and co-workers, this unique effect has been applied to many fields such as laser frequency stabilizations \cite{HS}, slow light and light storage\cite{LV,DF}. EIT is usually modeled by a three-level $\Lambda$ atomic system with a control light and a probe light coupling one upper level with two ground levels. Recently, it has shown that if a microwave field is applied as a perturbation to the two ground levels, EIT can be enhanced or suppressed by adjusting the relative phase of two light fields and the microwave field\cite{BL,HL}.
\begin{figure*}
\begin{center}
\onefigure{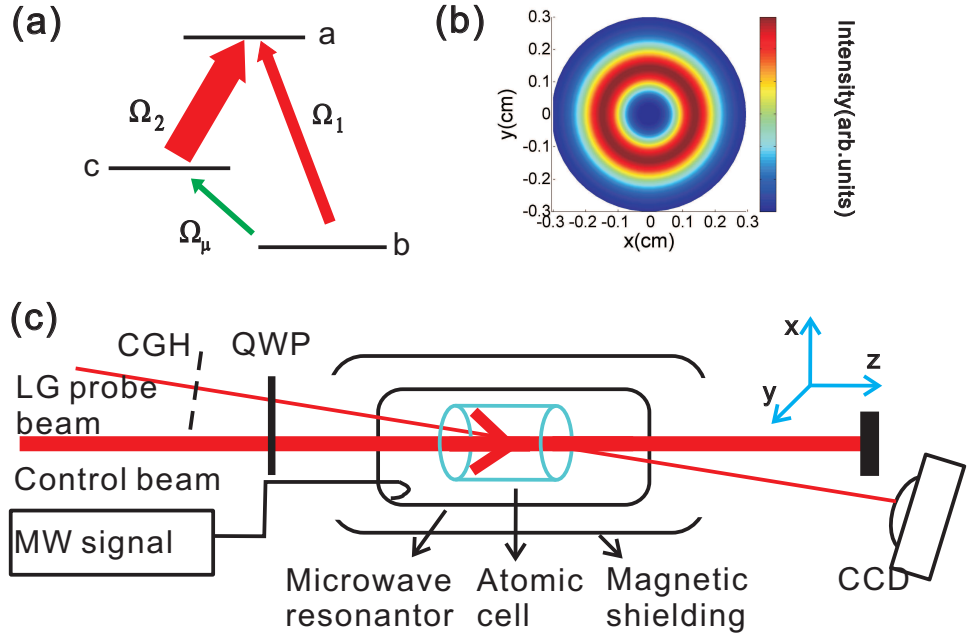}
\caption{(a) A three-level $\Lambda$ atomic system with two optical fields and one microwave field. The probe and microwave fields are detuned as $\Delta {\rm{ = }}{\omega _1} - {\omega _{ab}}{\rm{ = }}{\omega _\mu} - {\omega _{cb}}$. The control field resonates with the atomic transition. (b) A typical intensity distribution of the LG$_{0,l}$ beam, where $\l$ is a none zero integer. (c) Experimental scheme. The control and probe fields are phase-locked together. The small cross angle between the probe and control beams is for detecting probe beam conveniently. CGH is short for computer generated hologram. CCD is charge-coupled device. And QWP represents quarter wave plate.}
\label{fig1}
\end{center}
\end{figure*}

In this letter, we show that light carrying the OAM can be identified by using the EIT system which involves two optical fields and one microwave field. We demonstrate that when a beam with different LG modes propagates through such system, its transverse phase can be manipulated by the experimental configuration and eventually mapped into the intensity profile. The proposal provides an efficient method to distinguish the OAM states of LG modes based on atomic ensembles.

Our scheme starts from a three-level $\Lambda$ atomic system whose energy levels are shown in fig.~\ref{fig1}(a). A strong control beam $(E_2)$ and a weak probe beam $(E_1)$ are used to couple $\left| {\rm{c}} \right\rangle  \to \left| {\rm{a}} \right\rangle$ and $\left| {\rm{b}} \right\rangle  \to \left| {\rm{a}} \right\rangle$ atomic transitions, respectively. A microwave field $(E_\mu)$ is used to couple the two ground levels. This is a typical closed  $\Lambda $ system and has been intensively studied \cite{HL}. However, in the following analysis, we show that such system can be used to identify LG modes of the probe light. LG$_{p,l}$ mode has two parameters, $p$ and $l$, $l$ is the azimuthal index giving an OAM of $l\hbar$ per photon, and $p+1$ is the
number of radial nodes in the intensity distribution \cite{AM}. Typically, we set $p=0$ for simplicity. The intensity distribution of LG$_{0,l}$ beam shown in fig.~\ref{fig1}(b). Figure~\ref{fig1}(c) illustrates an experimental configuration to implement our proposal. A probe beam with LG mode produced by a computer generated hologram (CGH) incidents into the atomic cell with a small cross angle to the control beam. The atomic cell is placed inside a magnetic shielding cavity and a microwave resonator. In our scheme, the control and probe light are phase-locked to ensure the following coherent interaction process. The corresponding fields can be expressed as
\begin{eqnarray}
{{E_1} = {\varepsilon _1}{(r/w)^{\left| l \right|}}{e^{ - {{(r/w)}^2}}}\cos ({\omega _1}t - {k_1}z + {\phi _1} + l\varphi )},
\end{eqnarray}
\begin{eqnarray}
{{E_2} = {\varepsilon _2}\cos ({\omega _2}t - {k_2}z + {\phi _2})},
\end{eqnarray}
\begin{eqnarray}
{{E_\mu } = {\varepsilon _\mu }(z)\cos ({\omega _\mu }t + {\phi _\mu })},
\end{eqnarray}
where ${\varepsilon}$, ${\omega}$  and ${\phi}$  are the amplitude, angular frequency and initial phase, respectively. ${k_{1}} (k_{2})$ is the wave number for probe field (control field), $w$ is the waist of probe beam, and $l$ is the OAM number. For ${E_1}$ field, we have neglected the term of $\exp \left[ {\frac{{ik_0 r^2 z}}{{2(z^2  + z_r ^2 )}}} \right]\exp \left[ { - i(2p + \left| l \right|)\tan ^{ - 1} (\frac{z}{{z_R }})} \right]$, as this term has little influence to the result in this model.

To understand the probe beam propagation properties in such system, we need to know the atomic polarization seen by the probe beam. This can be obtained from the motion equation of density matrix, i.e.,
\begin{eqnarray}
\label{eq.1}
\dot \rho  =  - \frac{i}{\hbar }[{\rm{H, }}\rho ] - \frac{1}{2}\{ \Gamma ,\rho \},
\end{eqnarray}
where $H = \hbar {\omega _a}\left| a \right\rangle \left\langle a \right| + \hbar {\omega _b}\left| b \right\rangle \left\langle b \right| + \hbar {\omega _c}\left| c \right\rangle \left\langle c \right| + [{\wp _{ab}}{E_1}\left| a \right\rangle \left\langle b \right| + {\wp _{ac}}{E_2}\left| a \right\rangle \left\langle c \right| + {\wp _{bc}}{E_\mu }\left| c \right\rangle \left\langle b \right| + H.c.]$ is the Hamiltonian of the microwave modified EIT system.
${\wp _{ab}} = {\wp ^ * }_{ba} = e\left\langle {\rm{a}} \right|x\left| b \right\rangle$,
${\wp _{ac}} = {\wp ^ * }_{ca} = e\left\langle {\rm{a}} \right|x\left| c \right\rangle$ and
${\wp _{bc}} = {\wp ^ * }_{bc} = e\left\langle b \right|x\left| c \right\rangle $ are the matrix elements of the electric dipole moment. $\Gamma $ is the relaxation matrix. To solve eq.~(\ref{eq.1}), we introduce the slowly varying envelope as
${\rho _{ab}} = {\sigma _{ab}}{e^{(i{k_1}z - i{\omega _1}t - il\varphi )}}$, ${\rho _{ac}} = {\sigma _{ac}}{e^{(i{k_2}z - i{\omega _2}t)}}$, ${\rho _{cb}} = {\sigma _{cb}}{e^{(i\Delta kz - i({\omega _1} - {\omega _2})t)}}$, and $\rho _{\alpha \alpha }  = \sigma _{\alpha \alpha }$, where $\Delta k = {k_1} - {k_2}$\cite{MO}. Under the rotating wave approximation, eq.~(\ref{eq.1}) is changed to
\begin{eqnarray}
\label{eq.2}
{{\dot \sigma }_{ab}} = - {\Gamma _{ab}}{\sigma _{ab}} - i{\Omega _1}({\sigma _{aa}} - {\sigma _{bb}})
 + i{\Omega _2}{\sigma _{cb}}\notag\\- i{\Omega _\mu }{e^{i({\omega _1} - {\omega _2} - {\omega _\mu })t - i\Delta kz + il\varphi }}{\sigma _{ac}},
\end{eqnarray}
\begin{eqnarray}
\label{eq.3}
{{\dot \sigma }_{ac}} = - {\Gamma _{ac}}{\sigma _{ac}} - i{\Omega _2}({\sigma _{aa}} - {\sigma _{cc}}) + i{\Omega _1}{\sigma _{bc}}\notag\\
 - i{\Omega _\mu }^*{e^{ - i({\omega _1} - {\omega _2} - {\omega _\mu })t + i\Delta kz - il\varphi }}{\sigma _{ab}},
\end{eqnarray}
\begin{eqnarray}
\label{eq.4}
{{\dot \sigma }_{cb}} = - {\Gamma _{cb}}{\sigma _{cb}} - i{\Omega _\mu }{e^{i({\omega _1} - {\omega _2} - {\omega _\mu })t - i\Delta kz + il\varphi }}\notag\\\times({\sigma _{cc}} - {\sigma _{bb}})
 + i{\Omega _2}^*{\sigma _{ab}} - i{\Omega _1}{\sigma _{ca}},
\end{eqnarray}
where ${\Omega _1} = \frac{{{\wp _{ab}}{\varepsilon _1}}}{{2\hbar }}{(r/w)^{\left| l \right|}}{e^{ - {{(r/w)}^2}}}{e^{ - i{\phi _1}}}$, ${\Omega _2} = \frac{{{\wp _{ac}}{\varepsilon _2}}}{{2\hbar }}{e^{ - i{\phi _2}}}$, and
${\Omega _\mu } = \frac{{{\wp _{bc}}{\varepsilon _\mu }(z)}}{{2\hbar }}{e^{ - i{\phi _\mu }}}$ are the Rabi frequencies of the corresponding fields. At the steady state, assuming all atoms populate in the ground state $\left| {\rm{b}} \right\rangle $ due to the strong control field, i.e., $\rho _{{\rm{bb}}}  \approx 1$ and $\rho _{{\rm{aa}}}  = \rho _{{\rm{cc}}}  \approx 0$, we obtain an analytical solution of the atomic coherence term related to the probe beam,
\begin{eqnarray}
\label{eq.5}
{\sigma _{ab}} = \frac{{i{\Gamma _{cb}}{\Omega _1}}}{{{\Gamma _{ab}}{\Gamma _{cb}} + {{\left| {{\Omega _2}} \right|}^2}}} - \frac{{{\Omega _2}{\Omega _\mu }{e^{i( - \Delta kz + l\varphi )}}}}{{{\Gamma _{ab}}{\Gamma _{cb}} + {{\left| {{\Omega _2}} \right|}^2}}},
\end{eqnarray}
where ${\Gamma _{cb}} = {\gamma _{cb}} + i\Delta $ and ${\Gamma _{ab}} = {\gamma _{ab}} + i\Delta $. For realistic, we need to consider the motion of atoms such that $\Gamma$ should be modified. Without loss generality, we can neglect the Doppler width of two-photon transition and just take into account the Doppler width of one-photon transition. Thus the modified  $\Gamma$ can be expressed as,
${\Gamma _{ab}} = {\gamma _{ab}} + {k_1}u + i\Delta$, ${\Gamma _{ac}} = {\gamma _{ac}} + {k_2}u$  and ${\Gamma _{cb}} = {\gamma _{cb}} + i\Delta$ , where $u$ is the most probable speed of atoms\cite{HL,AJ}. Once ${\sigma _{ab}}$ is obtained, the atomic polarization governing the probe propagation can be determined, i.e.,
$P(z,t) = {\wp _{ab}}[{\rho _{ab}}(z,t) + c.c]$. From the Maxwell equations, we then can solve the propagation equation for the probe beam as:
\begin{eqnarray}
\label{eq.6}
\frac{{\partial {\Omega _1}}}{{\partial z}} = i\eta {\sigma _{ab}},
\end{eqnarray}
where $\eta  = \omega _1 N\wp _{ab} ^2 /(2\varepsilon _0 {\rm{c}}\hbar )$ is the coupling constant, $N$ is the atomic density, and ${\varepsilon _0}$ is the permittivity in vacuum. In this sense the propagation equation of probe beam with the LG mode passing through an EIT system controlled by a microwave field can be expressed as \cite{HL}:
\begin{eqnarray}
\label{eq.7}
\frac{{\partial {\Omega _1}}}{{\partial z}} =  - \frac{{{\eta}{\Gamma _{cb}}{\Omega _1}}}{{{\Gamma _{ab}}{\Gamma _{cb}} + {{\left| {{\Omega _2}} \right|}^2}}} - i\frac{{{\eta}{\Omega _2}{\Omega _\mu }{e^{ - i(\Delta kz - l\varphi )}}}}{{{\Gamma _{ab}}{\Gamma _{cb}} + {{\left| {{\Omega _2}} \right|}^2}}}.
\end{eqnarray}
The first term of right side of eq.~(\ref{eq.7}) generally describes a $\Lambda$-system EIT effect, while the second term is from the process involved the microwave field, which is an essential part to distinguish the LG modes in our scheme. Assuming the initial Rabi frequency of probe beam is $\Omega _{10} = \varepsilon _{10}{e^{ - {{(r/w)}^2}}}(r/w)^{\left| l \right|}$ at the position of $z_0$, the transmitted probe field passes through the atomic cell with length of $L$ can be expressed as:
\begin{eqnarray}
\label{eq.8}
{\Omega _1}({z_0} + L) = {\varepsilon _{10}}{e^{ - \alpha L}}{(\frac{r}{w})^{\left| l \right|}}{e^{ - {{(r/w)}^2}}} - [{e^{ - i\Delta k({z_0} + L)}}\notag\\
- {e^{ - i\Delta k{z_0} - \alpha L}}]\frac{{i\eta {\Omega _\mu }{\Omega _2}{e^{il\varphi }}}}{{({\Gamma _{cb}}{\Gamma _{ab}} + {{\left| {{\Omega _2}} \right|}^2})}}\frac{1}{{( - i\Delta k + \alpha )}},
\end{eqnarray}
where
$\alpha  = \eta \frac{{{\Gamma _{cb}}}}{{{\Gamma _{cb}}{\Gamma _{ab}} + {{\left| {{\Omega _2}} \right|}^2}}}$  is absorption coefficient. Thus the intensity profile can be calculated by
${\left| {{\Omega _1}({z_0} + L)} \right|^2}$.

\begin{figure}
\onefigure{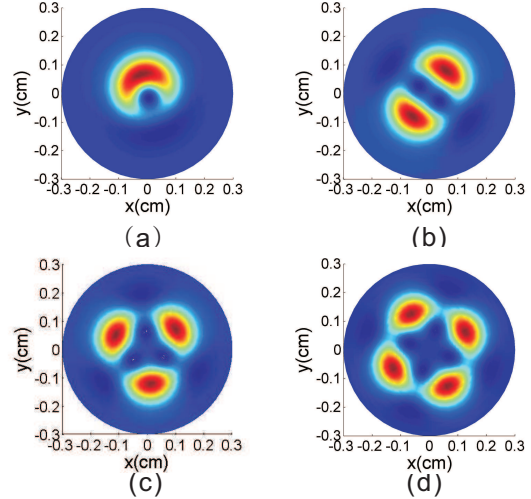}
\caption{The spatial intensity profiles of probe beam with the LG$_{0,l}$ mode behind the Rb cell. (a) to (d) correspond to $l$ from 1 to 4 and ${z_0}=3.3{\rm{cm}}$.}
\label{fig2}
\end{figure}

To show how our theory can be used to identify the LG modes, we use $^{87}$Rb D1 line (${5^2}{{\rm{S}}_{1/2}} \to {5^2}{{\rm{P}}_{1/2}}$, $\lambda {\rm{ = 795nm}}$) to numerically simulate the probe beam profile. In this case, EIT two ground levels are two hyperfine ground states of
${5^2}{{\rm{S}}_{1/2}}$ (F=1) and ${5^2}{{\rm{S}}_{1/2}}$ (F=2). The corresponding parameters are set as
${\gamma_{ab}}=6$,
${\gamma_{bc}}={10^{ - 3}}$,
${\varepsilon_{10}}=0.1$,
${\Omega_2}=1$,
${\Omega_\mu}=0.02$,
$\eta=0.9$,
$L=3{\rm{cm}}$,
${k_1}u=500$,
$\Delta k=1.43{\rm{c}}{{\rm{m}}^{-1}}$, and
$w=1{\rm{mm}}$. The above values are common in an EIT system and can be easily obtained in the experiment. Figure~\ref{fig2} shows the intensity distributions of probe beam behind the Rb cell for the different OAM numbers. Clearly, the transverse phase information carried by the LG mode now is converted to the intensity distribution, and different OAM number $l$ corresponds to the different bright area number. By counting the number of bright areas, an unknown LG mode can be easily recognized.
\begin{figure}
\onefigure{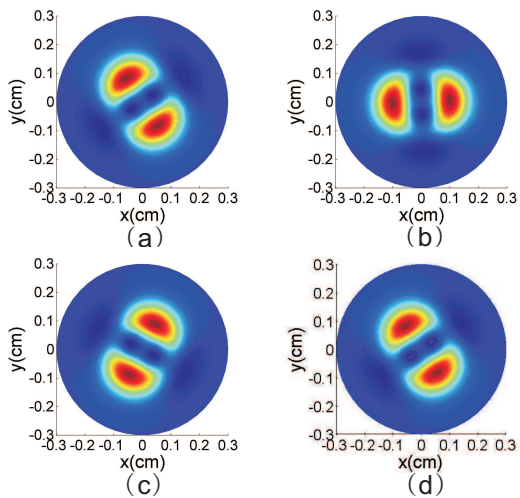}
\caption{The spatial intensity profiles of probe beam with the LG$_{0,2}$ mode versus the different positions of the Rb cell. (a) to (d) correspond to ${z_0}$ = 5, 6.46, 7.93, and $9.4{\rm{cm}}$ respectively.}
\label{fig3}
\end{figure}

As we know, OAM has both negative and positive values. For these mode beam identification, it is usually difficult to sort in other ways \cite{AF}. The intensity distributions shown in fig.~\ref{fig2} do not provide any additional information about the sign of OAM. However, as the phase of control beam and probe beam is locked in our configuration, these two beams form a wave packet along $z$ axis. The frequency of the wave packet equals to the frequency difference of two beams. So the total relative phase ($\Delta kz - l\varphi $) will change linearly with the position of the Rb cell moving along $z$ axis. In such way, OAM states of $+l$ and $-l$ can also be distinguished in our scheme. Fig.~\ref{fig3}
shows the intensity profile variations of a LG$_{0,2}$ probe beam with the positions of the Rb cell moving along $z$ direction. We can see that the total relative phase changing at the different Rb cell positions arouses a rotation of intensity profile. For the typical $^{87}$Rb D1 line, the frequency difference of two ground hyperfine levels is 6.83GHz, which corresponds to the beat wavelength $L'=4.4cm$. Therefore the intensity profile is recovered as the Rb cell moves one beat wavelength as shown in fig.~\ref{fig3}. Moreover, our theoretical analysis indicates that the rotating direction of intensity pattern is related to the sign of the OAM. In other words, the reverse rotating direction of intensity profile indicates that the LG mode carries an opposite OAM, i.e., LG$_{0,-2}$. Thus we can facilely sort the positive and negative LG modes by moving the Rb cell and seeing the rotating directions of the intensity profile. Furthermore, we use the absorption to detect the rotation direction of the phase in the LG beam, which is related to the atomic cell length and beat wavelength. From eq.~(\ref{eq.8}), a simple derivation shows that the spatial profile will have a good distinguishability when $L\neq nL'$, and $n$ is an integer.

\begin{figure}[!h]
\onefigure{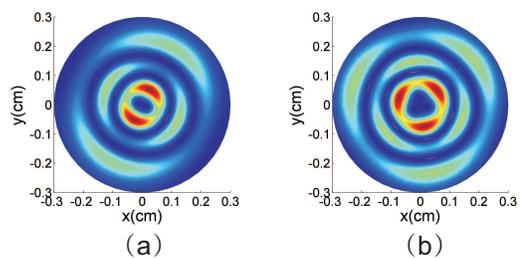}
\caption{The spatial intensity profiles of probe beam with the LG$_{2,l}$ mode behind the Rb vapor cell. (a) and (b) correspond to $l$ is 2 and 3, respectively.}
\label{fig4}
\end{figure}
Our scheme is also valid for LG$_{p,l}$ mode with $p\neq0$. We calculate the LG$_{2,l}$ mode as an example and illustrate the results in fig.~\ref{fig4}. Obviously, three rings appear corresponding to the radial mode of the LG beam. And each ring has a specific bright area number related to its OAM number. Thus the general LG$_{p,l}$ mode can also be identified. Interestingly, the intensity distribution is complementary between the adjacent rings. This is due to the intrinsic transverse phase distribution of the LG mode.

It should be mentioned that diffraction and diffusion effects are not considered in our scheme. These effects are important in light storage experiment. However in our scheme, the probe light propagates through the atomic cell together with the control light and the cell length is usually short, so the above two effects should not play an important role. In this letter, we assume the phase and amplitude of microwave are constant. It is also interesting to note that a similar rotation occurs by changing the phase of microwave. Meanwhile, the strength of microwave would affect the distinguishability of intensity pattern.

Before drawing a conclusion, we should address that our method for mode sorting is not as convenient as the other optical methods. However, we exploit an OAM sorting way based on the prospective atomic assembles. It may be useful for quantum repeater and higher-dimensional quantum communication. Furthermore, this method can be used for reducing the effect of diffusion on the visibility of the reconstructed image in the light storage.

In summary, we have theoretically exhibited how to recognize the LG mode based on atomic ensembles. The microwave is used to perturb the atomic coherence of two ground levels. As the transverse phase of probe beam variation, the phase information is then converted to the transmission intensity change. So different order of the LG modes can be distinguished by the transverse intensity distribution. Moreover we find that our scheme can be easily adapted to sorting the negative and positive modes of the LG beam by moving the position of the atomic cell. This novel approach to sort the LG mode may provides potential applications in quantum memory and light storage with the EIT system. It may also be used in the phase contrast imaging with high modulation depth or in phase-to-amplitude conversion in an atomic system.
\acknowledgments
This work is financially supported by the National Natural Science Foundation of China under grants 11074198 and 11004158, the Special Prophase Project on the National Basic Research Program of China under grant 2011CB311807, and the Fundamental Research Funds for the Central Universities.

\end{document}